\documentclass[doublecol,figures]{epl2}

\title{ Noise induced forced synchronization of
global variables in coupled bistable systems}
\shorttitle{Title} 

\author{Jos\'e M. Casado, Jos\'e G\'omez-Ord\'o\~nez and Manuel
  Morillo }

\institute{
  Universidad de Sevilla.
F\'{\i}sica Te\'orica. Apartado de Correos 1065. Sevilla 41080. Spain}


\abstract{ We analyze the noise induced synchronization between a
collective variable characterizing a complex system with a finite
number of interacting bistable units and time periodic driving
forces. A random phase process associated to the collective
stochastic variable is defined. Its average phase frequency and
average phase diffusion are used to characterize the phenomenon. Our
analysis is based on numerical solutions of the corresponding set of
Langevin equations. }

\begin{document}

\maketitle

\section{Introduction}
Synchronization phenomena in nonlinear systems have been widely studied
in very different contexts \cite{pikovskybook2001,chaos2003,ojalvo}. In
complex nonlinear autonomous systems with many interacting subunits,
nonlinear dynamics might lead to self-synchronization between the
individual elements. Noisy environments and externally applied time
periodic forces might also influence synchronization. One might be
interested on how noise and driving forces affect the synchronization
between the different subunits, or the focus of interest might be the
noise induced forced synchronization  to the external driving force of a
suitably defined single collective variable characterizing the system as
a whole.

In the case of systems characterized by just a single noisy
dynamical variable subject to the action of a deterministic external
time-periodic signal, noise-induced forced synchronization has been
amply studied. With an appropriate definition of a random phase
process associated to the response of the system, theories of
noise-induced forced synchronization have been developed and
analytical expressions for the quantities characterizing the forced
synchronization mechanism have been obtained
\cite{freund1,callenbach,freund2,casado,talkner,talkner1}. The
theory of forced synchronization in a single variable system has
also been extended to the quantum regime of incoherent tunneling
transitions in driven dissipative systems \cite{goychuk}.
Synchronization effects in complex systems formed by arrays of
excitable systems and stochastic resonators have been analyzed in
\cite{ojalvo,neiman}.

In this work, we consider a model describing a
\emph{finite} set of $N$ interacting \emph{bistable} subsystems,
each of them characterized by a single degree of freedom $x_i$
($i=1,\ldots,N$), whose dynamics is governed by the Langevin
equations \cite{deszwa,josem}
\begin{equation}
\dot{x}_i=x_i-x_i^3+\frac{\theta}{N}\sum_{j=1}^N(x_j-x_i)+\xi_i(t)+F(t).
\label{modeleq1}
\end{equation}

Here, the $\xi_i(t)$'s are Gaussian white noises with zero average and
$\langle \xi_i(t)\xi_j(s)\rangle=2D\delta_{ij}\delta(t-s)$, $\theta$
is the parameter defining the strength of the interaction between
subsystems, and $F(t)=F(t+T)$ is an external driving force of period
$T$. The amplitude of the driving force is supposed to be large so
that a linear response approximation of the dynamics is not
adequate. Nonetheless, we are interested in noise induced
synchronization, so the driving amplitude is subthreshold in the
sense that, in the absence of noise, the external driving by itself
can not induce the phenomenon.

We focus our interest on a single global variable, rather than on the
variable characterizing an individual subsystem. In a recent work, we
have analyzed the phenomenon of Stochastic Resonance associated to the
global variable of the same system \cite{josem,cubero}. In this work, we
analyze the phase process associated to the noise induced jumps of the
global variable between its dynamical attractors. Noise induced forced
synchronization between the global variable and the driving force is
characterized by the behavior of the average phase frequency and
the average phase diffusion constant with the noise strength.

\section{The random phase process, its average phase frequency and the
  diffusion constant}

The variable of interest is the collective output $S(t)$ defined as
\begin{equation}
\label{collec} S(t)= \frac 1N \sum_{i=1}^N  x_i(t).
\end{equation}

In the limit $N \rightarrow \infty$, an asymptotically valid Langevin
equation for $S(t)$ can be obtained from Eq.  (\ref{modeleq1})
\cite{mor95}.  For finite systems the validity of this Langevin equation
is questionable and we rely on the solution of the Langevin equation for
each $x_i$ to obtain information about the global process $S(t)$. For
low noise strengths and driving forces with sufficiently large periods,
our numerical simulations show that a random trajectory of $S(t)$
contains essentially small fluctuations around two values (attractors)
and random, sporadic transitions between them. For each realization of
the noise term, we then introduce a random phase process, $\phi(t)$,
associated to the stochastic variable $S(t)$ as follows.  We refer to a
``jump'' of $S(t)$ along a trajectory, when a very large fluctuation
takes the $S(t)$ trajectory from a value near an attractor to a value in
the neighborhood of the other attractor.  We count $N^{(\alpha)}(t)$,
the number of jumps in the $\alpha$ trajectory of the process $S(t)$
within the interval $(0,t]$ . A trajectory of the phase process is then
constructed as
\begin{equation}
\phi^{(\alpha)}(t)=\pi N^{(\alpha)}(t),
\end{equation}
so that $\phi(t)$ increases by $2\pi$ after every two consecutive
jumps.

 The first two moments of the phase process are estimated as
\begin{equation}
\label{phaseaverage} \left\langle \phi(t)\right\rangle
=\frac{1}{\mathcal M} \sum_{\alpha=1}^{\mathcal M}
\phi^{(\alpha)}(t)
\end{equation}
\begin{eqnarray}
\label{phasedispersion}
v(t)&=&\left\langle
\left[\phi(t)\right]^2\right\rangle -\left\langle
\phi(t)\right\rangle^2 \nonumber \\
&=&\frac{1}{\mathcal M}
\sum_{\alpha=1}^{\mathcal M}
\left[\phi^{(\alpha)}(t)\right]^2-\frac{1}{\mathcal M^2}
\left[\sum_{\alpha=1}^{\mathcal M} \phi^{(\alpha)}(t)\right]^2
\end{eqnarray}
where, ${\mathcal M}$ is the number of generated random trajectories
(typically, 3000 trajectories for the results presented in this work).

The instantaneous phase frequency is easily determined from the time
derivative of $\left\langle \phi(t)\right\rangle $.  After a
sufficiently large number of periods of the driving force, $n$, the
system forgets its initial preparation, but the instantaneous phase
frequency is still a function of time. Then, we define a cycle
average phase frequency $\overline{\Omega}_\mathrm{ph}$ by averaging
the instantaneous phase frequency over a period of the external
driving, \cite {casado,goychuk}

\begin{equation}
\label{frequency2} \overline{\Omega}_\mathrm{ph}=\frac 1T
\int_{nT}^{(n+1)T} dt\,\frac {d\langle \phi(t) \rangle}{dt}=
\frac{\left\langle
\phi\left[(n+1)T\right]\right\rangle-\left\langle
\phi(nT)\right\rangle}{T}
\end{equation}

Similarly, the cycle average phase diffusion coefficient is evaluated
from the instantaneous slope of the variance $v(t)$ as
\cite {casado,goychuk},
\begin{equation}
\label{diffusion2}
\overline{D}_\mathrm{ph}=\frac 1T
\int_{nT}^{(n+1)T} dt\,\frac {d\langle v(t) \rangle}{dt}=
\frac{v\left[(n+1)T\right]
  -v\left(nT\right)}{T}
\end{equation}
In previous works, approximate anlytical expressions for these two
quantities have been derived for the $N=1$ problem in the classical
\cite{freund1,casado,talkner1,prager} and quantum cases
\cite{goychuk}. Those expressions can not be applied to the
collective variable of an $N$-particle problem, as a closed Langevin
or Fokker-Planck equation for $S(t)$ for a finite size system does
not exist. Thus, we will rely on numerical solutions of the Langevin
equations, Eq. (\ref{modeleq1}), for the estimation of the average
phase frequency and diffusion coefficients in Eqs.
(\ref{frequency2}) and (\ref{diffusion2}). The numerical method used
to solve the Langevin dynamics has been detailed in \cite{casgom03}.

\section{Results}
We have analyzed the forced syncronization phenomenon for the collective
variable $S(t)$ for several values of the
coupling strength, $\theta$, the noise strength, $D$, and two types of
periodic forces: sinusoidal forces ($F(t)=A\cos \Omega t$) and
rectangular forces  ($F(t)=A$ ($F(t)=-A$) if $t\in [n T/2, (n+1)T/2)$
with $n$ even (odd)).

\begin{figure}
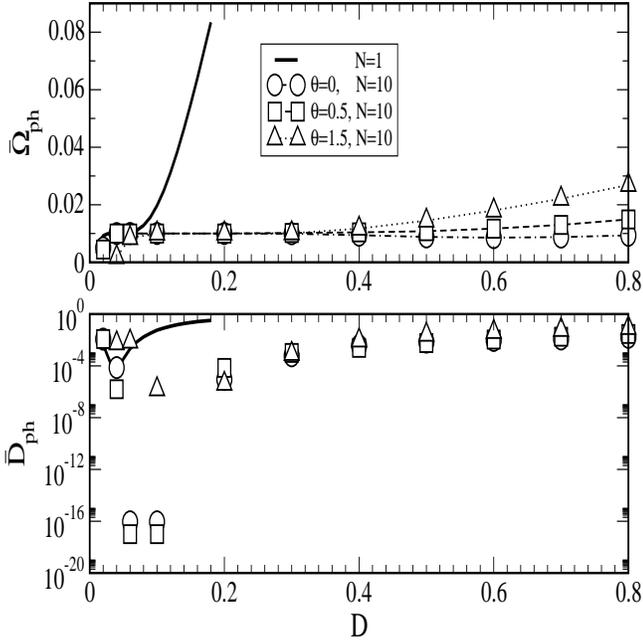

\onefigure[width=8.5cm,height=8.5cm]{Fig1.eps} \caption{ The
behavior of the average phase frequency,
$\overline{\Omega}_\mathrm{ph}$, and the phase
    diffusion constant, $\overline{D}_\mathrm{ph}$, with the noise strength, $D$, for
    a sinusoidal driving term with  $A=0.3$ and $\Omega=0.01$ and several values of the interaction strength $\theta$.}
\label{fig.1}
\end{figure}

In Fig. \ref{fig.1} we depict the behavior of the average phase
frequency, $\overline{\Omega}_\mathrm{ph}$, and diffusion constant,
$\overline{D}_\mathrm{ph}$, obtained from numerical simulations. The
external driving force is taken to be sinusoidal with $A=0.3$ and
$\Omega=0.01$. In the absence of noise, the variable $S(t)$ does not
jump between the attractors. Thus, for these parameter values, the
external driving is subthreshold and we have noise induced effects.  It
is clear from the picture that, even for a single subsystem, $N=1$,
forced synchronization exists as there is a range of noise values for
which $\overline{\Omega}_\mathrm{ph}$ matches the external frequency,
$\Omega$ and the diffusion constant is rather small, with a minimum
around $D\approx 0.02$. As we discused previously in \cite{casado},
these numerical results agree very well with the analytical expressions
reported in \cite{casado}. For a set of $N=10$ independent subsystems
($\theta=0$), forced synchronization for the collective variable is very
much enhanced with respect to that observed in a single particle
variable. The matching between the driving frequency and the average
phase frequency extends over a very large range of noise values, for
which the corresponding diffusion constant is very much reduced. It
should be pointed out that, in this and subsequent figures, the fact
that a ``range'' of $D$ values shows up in the plots, for which
$\overline{D}_\mathrm{ph}$ seems to be constant, is due to the lack of
machine precision to compute very small numbers. It is safe to
say that for the noise values in those ranges, the phase diffusion
constant is zero, and that the collective variable jumps every
half-period of the driver without skipping any change in sign of the
applied force.  

\begin{figure}
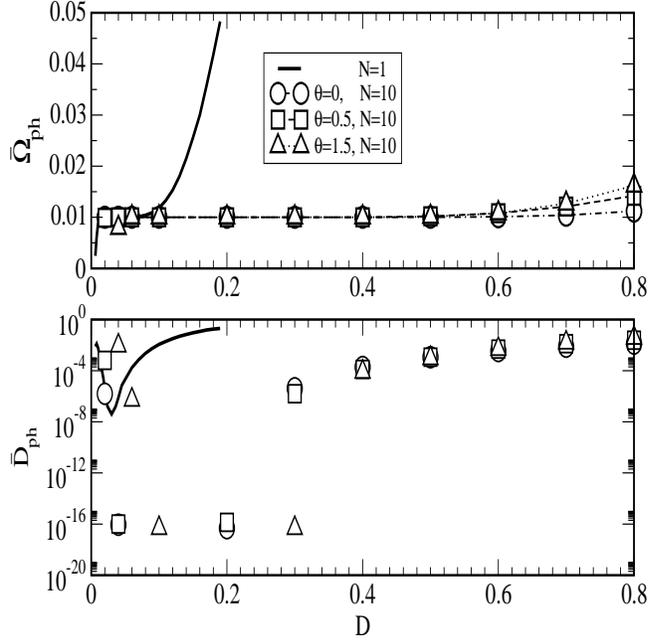

\onefigure[width=8.5cm,height=8.5cm]{Fig2.eps} \caption{ The behavior
of the average phase frequency, $\overline{\Omega}_\mathrm{ph}$, and
the phase
    diffusion constant, $\overline{D}_\mathrm{ph}$, with the noise strength, $D$, for
    a rectangular driving term with  $A=0.3$ and fundamental frequency
    $\Omega=0.01$ and several values of the interaction strength $\theta$.}
\label{fig.2}
\end{figure}

The results for a rectangular input force with an amplitude $A=0.3$
and fundamental frequency $\Omega=0.01$ are shown in Fig.
\ref{fig.2}. The behaviors of $\overline{\Omega}_\mathrm{ph}$ and
$\overline{D}_\mathrm{ph}$ with the noise strength for the different
sets of parameter values is qualitatively similar to the one
observed in Fig.  \ref{fig.1} for a sinusoidal driving.
Quantitatively, for a rectangular driving, the noise induced forced
synchronization is enhanced with respect to that observed with a
sinusoidal input with the same amplitude and fundamental frequency, as
indicated by the wider range of $D$ values for which frequencies match.

\begin{figure}
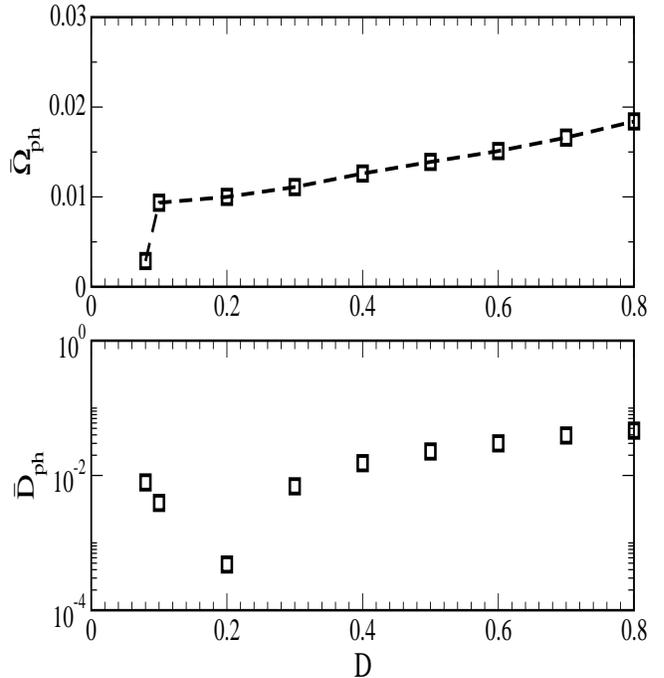

\onefigure[width=8.5cm,height=9cm]{Fig3.eps} \caption{ The behavior
of the average phase frequency, $\overline{\Omega}_\mathrm{ph}$, and
the phase
    diffusion constant, $\overline{D}_\mathrm{ph}$, with the noise strength, $D$, for
    a sinusoidal driving term with  $A=0.1$ and $\Omega=0.01$.}
\label{fig.3}
\end{figure}

We have analyzed the influence of increasing the driving fundamental
frequency or decreasing the external amplitude. In Fig. \ref{fig.3}
we depict the results obtained for $\overline{\Omega}_\mathrm{ph}$
and $\overline{D}_\mathrm{ph}$ for a rectangular input with $A=0.1$
and $\Omega=0.01$, with $N=10$ interacting subsystems with $\theta =
0.5$. Comparison with the corresponding results in Fig. \ref{fig.1}
indicates that synchronization for this smaller amplitude driving is
much weaker than for the larger amplitude case. Not only the noise
range for which the average phase frequency matches the external
frequency is reduced, but the dip in the phase diffusion constant
plot is much less pronounced.

In the next figure, we plot the results of our simulations for a
rectangular input with $A=0.3$ (the same as in Fig. \ref{fig.1}) and
$\Omega=0.1$, with $N=10$ interacting subsystems with $\theta =
0.5$. The increase in the fundamental driving frequency with respect to
the value used in  Fig. \ref{fig.1} leads also to a reduction of the noise
range for which noise induced forced synchronization is
optimal. Nonetheless, the dip in the $\overline{D}_\mathrm{ph}$ vs. the
noise strength seems to be of the same order of magnitude as observed
for the lower frequency case depicted in Fig. \ref{fig.1}.

\begin{figure}
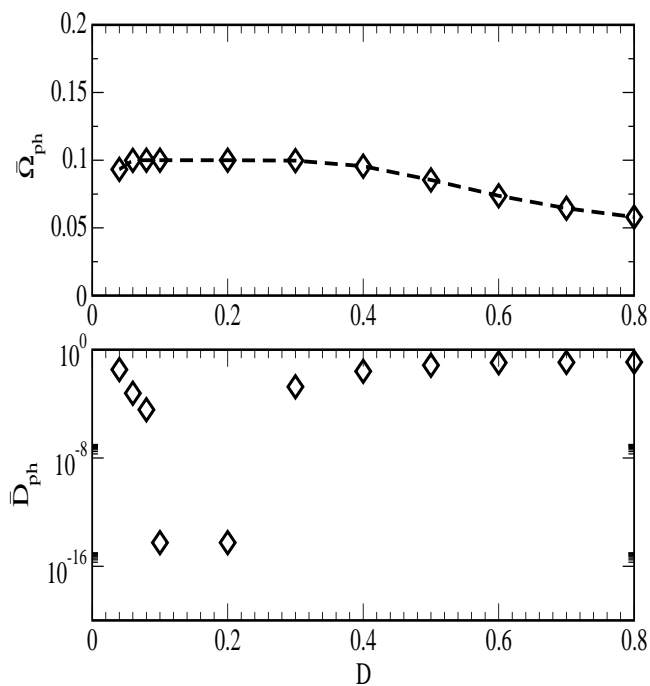

\onefigure[width=8.5cm,height=9cm]{Fig4.eps} \caption{ The behavior
of the average phase frequency, $\overline{\Omega}_\mathrm{ph}$, and
the phase diffusion constant, $\overline{D}_\mathrm{ph}$, with the
noise strength, $D$, for a rectangular driving term with $A=0.3$ and
$\Omega=0.1$.} \label{fig.4}
\end{figure}

The weakening of noise induced forced synchronization as the driving
amplitude and/or the driving period decrease is to be expected.
Noise induced forced synchronization is due to the modification by
the applied force of the intrinsic noisy dynamics of the set of
bistable systems. Noise itself makes the collective variable jumps
between attractors with a certain probability distribution. The
external driving greatly modifies this distribution by forcing the
jumps to be coherent with the periodicity of the external driving.
The external force tilts the potential surface such that every
half-period, jumps towards one attractor are much favored against
the ones towards the other one. Thus, stronger amplitudes and longer
driving periods, provide potential surfaces which remain tilted
during a time interval long enough for the noise induced jumps of
the subsystems variables to happen almost simultaneously every
half-period of the driver.

In conclusion, we have numerically demonstrated the phenomenon of
noise induced forced synchronization in finite arrays of bistable
systems. Interactions between the subunits do not destroy this
coherent effect. Indeed, the collective synchronization is much
enhanced with respect to the one observed for a single bistable
unit. Driving forces with short periods and small amplitudes tend to
weaken drastically the phenomenon.

 \acknowledgments We acknowledge the support
of the Ministerio de Educaci\'on y
  Ciencia of Spain (FIS2005-02884) and the Junta de Andaluc\'{\i}a.

\end{document}